\documentstyle[aaspp4,tighten,flushrt]{article}

\newcommand{\beq}{\begin{equation}}
\newcommand{\eeq}{\end{equation}}
\newcommand{\beqa}{\begin{eqnarray}}
\newcommand{\eeqa}{\end{eqnarray}}

\def\d{\delta}

\font\BF=cmmib10

\def\k{{\hbox{\BF k}}}
\def\x{{\hbox{\BF x}}}

\def\la{\mathrel{\mathpalette\fun <}}
\def\ga{\mathrel{\mathpalette\fun >}}
\def\fun#1#2{\lower3.6pt\vbox{\baselineskip0pt\lineskip.9pt
\ialign{$\mathsurround=0pt#1\hfill##\hfil$\crcr#2\crcr\sim\crcr}}}

\input epsf
\begin{document}
\hfill{\small FERMILAB-Pub-99/xxx-A}

\title{The Projected Three-point Correlation Function: Theory and 
Observations}

%
%
\author{Joshua A. Frieman$^{1,2}$ and Enrique Gazta\~naga$^{3}$}

\vskip 1pc

\affil{${}^1$ NASA/Fermilab Astrophysics Center, Fermi National 
Accelerator Laboratory, \\ P.O. Box 500, Batavia, IL  60510}

\affil{${}^2$ Department of Astronomy and Astrophysics, 
University of Chicago, Chicago, IL 60637}


\affil{${}^3$ Institut d'Estudis Espacials de Catalunya, IEEC/CSIC,
\\ Edf. Nexus-104 - c/ Gran Capitan 2-4, 08034 Barcelona, 
Spain}

%
\begin{abstract}
%
We report results for the angular three-point galaxy correlation 
function in the APM survey and compare them with theoretical 
expectations. For the first time, these measurements 
extend to sufficiently large scales to probe the weakly non-linear 
regime. On large scales, the results are in good agreement with the 
predictions of non-linear cosmological perturbation theory,  
for a model with initially Gaussian fluctuations 
and linear power spectrum $P(k)$ consistent with 
that inferred from the APM survey. 
These results reinforce the conclusion that 
large-scale structure is driven by non-linear gravitational 
instability and that APM galaxies are relatively 
unbiased tracers of the mass on large scales; they also 
provide stringent constraints upon 
models with non-Gaussian initial conditions and strongly 
exclude the standard cold dark matter model. 

\end{abstract}
\clearpage 

%
%
\section{Introduction}
\label{intro}
%
%

Traditionally, two-point statistics, the auto-correlation function
$\xi(r)$ and the power 
spectrum $P(k)$, have been the dominant benchmarks for testing theories of  
structure formation. However, with the advent of large galaxy 
surveys and the development of non-linear cosmological perturbation 
theory and large N-body simulations, it has become clear that higher-order 
correlations 
provide new probes of large-scale structure. In particular,
the $N>2$-point functions and moments test models for bias---the 
relation between the galaxy and mass distributions---and constrain 
non-Gaussianity in the initial conditions (Fry \& Gazta\~naga 1993, 
Frieman \& Gazta\~naga 1994, Gazta\~naga 1994, Gazta\~naga \& 
Frieman 1994, Fry 1994, Fry \& Scherrer 1994, Jaffe 1994, 
Fry, Melott, \& Shandarin 1995, 
Juszkiewicz, et al. 1995, Chodorowski \& Bouchet 1996, Gazta\~naga 
\& Mahonen 1996, Szapudi, Meiksin, \& Nichol 1996, 
Jing 1997, Matarrese et al. 1997, Mo, Jing, \& White 1997, 
Verde et al. 1998, Scoccimarro et al. 1998,
Scoccimarro, Couchman, \& Frieman 1998). 

The galaxy three-point function $\zeta$ has been measured in 
several angular and redshift catalogs 
(Peebles \& Groth 1975, Peebles 1975, Groth \& Peebles 1977, 
Fry and Seldner 1982, Bean et al. 1983, Efstathiou \& Jedredjewski 1984, 
Hale-Sutton et al. 1989, Baumgart \& Fry 1991, Jing, Mo, \& Boerner 1991, 
Jing \& Boerner 1998).
Given the limited volumes covered by these 
surveys, these measurements generally 
probed the three-point function on scales $0.1 \la r \la 10 ~h^{-1}$ Mpc.
They established that galaxies cluster hierarchically: on these 
scales, the hierarchical three-point amplitude, defined by 
\beq
Q(x_{12},x_{13},x_{23}) \equiv {\zeta(\x_{1},\x_{2},\x_{3}) \over 
\xi(x_{12})\xi(x_{13})+\xi(x_{12})\xi(x_{23})+\xi(x_{13})\xi(x_{23})}~,
\label{Q3def}
\eeq
(with $x_{ij}=|\x_i-\x_j|$) 
is nearly constant, independent of scale and configuration, with 
values in the range $Q \simeq 0.6-1.3$, depending on the catalog. 
This hierarchical form is consistent with expectations from 
N-body simulations in the non-linear regime (e.g., Matsubara \& Suto 1994, 
Fry, Melott, \& Shandarin 1993, Scoccimarro et al. 1998, 
Scoccimarro \& Frieman 1998).

On larger scales, $r \ga 10 ~h^{-1}$ Mpc, in the weakly 
non-linear regime where $\xi(r) \la 1$, 
non-linear perturbation theory (PT)---corroborated by 
N-body simulations---predicts 
that $Q$ becomes strongly dependent on the {\it shape} of the 
triangle defined by the three points $\x_i$ 
(Fry 1984, 
Fry, Melott, \& Shandarin 1993, Jing \& Boerner 1997, Scoccimarro 1997, 
Gazta\~naga \& Bernardeau 1998, Scoccimarro et al. 1998).
The large-scale configuration-dependence of the three-point function 
is characteristic of the non-linear 
dynamics of gravitational instability and is 
sensitive to the initial power spectrum, to the bias, and to 
non-Gaussianity in the initial conditions. Thus measurement of the 
three-point function and its shape dependence should provide 
a powerful probe of structure formation models. 
Measurements of higher-order galaxy correlations on large scales 
have been confined so far to volume-averaged correlation functions---the 
one-point cumulants $S_N$---and their cousins (e.g., cumulant correlators). 
These statistics, based on counts-in-cells, are computationally easier to 
measure than the corresponding $N-$point functions and, being 
averages over many $N-$point configurations, can be measured 
with higher signal to noise as well. On the other hand, 
this averaging process by definition destroys the information 
about configuration-dependence contained in the $N-$point functions 
themselves. 


In this Letter, 
we report measurement of the projected three-point function on large scales 
in the APM Galaxy Survey (Maddox et al. 1990), compare the results with   
predictions of non-linear perturbation theory (PT) and N-body simulations, 
and briefly discuss their implications for bias and non-Gaussianity.



We first recall the perturbative expression for the three-point function.
To lowest   
order, the statistical properties of the density contrast field, 
$\delta(\x) = (\delta \rho(\x)/{\bar \rho})-1$, 
are characterized by the auto-correlation function $\xi(x_{12}) = \langle 
\delta(\x_1)\delta(\x_2)\rangle$, and its Fourier transform, the 
power spectrum $P(k)$, 
where 
$\langle \delta(\k_1)\delta(\k_2)\rangle = (2\pi)^3 P(k_1) 
\delta_{\rm D}^3(\k_1+\k_2)$; explicitly,  
\beq
\xi(r) = {1\over 2\pi^2}\int dk k^2 P(k){\sin(kr)\over kr}~.
\label{xidef}
\eeq
In the results to be shown below, we consider the cold dark matter (CDM) 
family of models, with shape parameter $\Gamma = \Omega h$ ranging 
from 0.5 (standard CDM) to 0.2. We also consider a model consistent with 
the linear 
power spectrum inferred from the APM survey itself (Baugh \& Gazta\~naga 
1996),
\beq
P_{\rm APM-like}(k) = {A k\over \left[1+(k/0.05 )^2\right]^{1.6}}~,
\label{Pkapm}
\eeq
with $k$ in $h~ {\rm Mpc}^{-1}$.
For Gaussian initial conditions, to linear order in $\delta$, the 
connected $N>2$-point functions vanish. In second-order perturbation 
theory, the three-point correlation function $\zeta(\x_1,\x_2,\x_3) \equiv 
\langle \delta(\x_1)\delta(\x_2)\delta(\x_3)\rangle$ can be expressed 
as (Fry 1984, Jing \& Boerner 1997, Gazta\~naga \& Bernardeau 1998)
\beqa
\zeta(\x_1,\x_2,\x_3) &=& {10\over 7}\xi(x_{12})\xi(x_{13})+{4\over 7}\left[
{{\Phi'}(x_{12}){\Phi'}(x_{13})\over x_{12}x_{13}} + 
\left(\xi(x_{12})+2{{\Phi'}(x_{12})\over x_{12}}\right)\left(
\xi(x_{13})+2{{\Phi'}(x_{13})\over x_{13}}\right)\right] \nonumber \\
&-& {\x_{12}\cdot \x_{13}\over x_{12}x_{13}} \left({\xi'}(x_{12})
{\Phi'}(x_{13})+{\xi'}(x_{13}){\Phi'}(x_{12})\right) \nonumber \\
&+& {4\over 7}\left({\x_{12}\cdot \x_{13}\over x_{12}x_{13}}\right)^2
\left(\xi(x_{12})+3{{\Phi'}(x_{12})\over x_{12}}\right)\left(
\xi(x_{13})+3{{\Phi'}(x_{13})\over x_{13}}\right)+{\rm cyclic ~permutations}, 
\label{zeta1}
\nonumber \\
\eeqa
where 
\beq
\Phi(r) \equiv {1\over 2\pi^2}\int dk  P(k){\sin(kr)\over kr}~,
\label{phidef}
\eeq
and $f'(x)=df/dx$. In eqns.(\ref{xidef})-(\ref{phidef}), the power 
spectrum and auto-correlation function are implicitly evaluated at linear 
order in perturbation theory;  
we expect the leading-order result  
(\ref{zeta1}) to be valid in the weakly non-linear regime ($\xi \ll 1$) but 
to break down for $\xi(r) \ga 0.5$.
Eqn.(\ref{zeta1}) was derived for the 
Einstein-de Sitter ($\Omega_m=1$) model, but it is known to 
be an excellent approximation for $\Omega \ga 0.1$ (Bouchet et al. 1992, 
1995, Bernardeau 1994, Catelan et al. 1995, Scoccimarro et al. 1998, 
Kamionkowski \& Buchalter 1998).
For non-Gaussian initial conditions, there will in general be 
corrections to (\ref{zeta1}) which depend on the initial (linear) 
three- and four-point functions (Fry \& Scherrer 1994). 

The expressions above apply to unbiased tracers of the 
density field; since galaxies of different types (e.g., 
ellipticals and spirals) have different clustering properties, we know 
that at least some galaxy species are biased. As an example, suppose
the probability of forming a luminous galaxy depends  
only on the underlying mean density field in its immediate vicinity. 
Under this simplifying assumption, the relation between the galaxy
density field $\d_g(\x)$ and the mass density field $\d(\x)$ is  
$\d_g(\x) = f(\d(\x)) = \Sigma_n b_n \d^n$, where $b_n$ are the
bias parameters. To leading order in perturbation theory, this local bias 
scheme implies 
\beq
\xi_g(x)=b_1^2 \xi(x)~,
\label{xig}
\eeq
\beq
\zeta_g(\x_1,\x_2,\x_3) = b_1^3 \zeta(\x_1,\x_2,\x_3)+b^2_1b_2
[\xi(x_{12})\xi(x_{13})+\xi(x_{12})\xi(x_{23})+\xi(x_{13})\xi(x_{23})]~,
\label{Zg}
\eeq
and therefore 
(Fry \& Gazta\~{n}aga 1993, Fry 1994) 
\beq
Q_{g} = {1 \over b_1} Q_{\delta} + {b_2 \over b^2_1} ~.
\label{Qg}
\eeq
Gazta\~{n}aga \& Frieman (1994) have used the corresponding 
relation for the skewness $S_3$ to infer $b_1 \simeq 1$, 
$b_2 \simeq 0$ from the APM catalog, but the results are 
degenerate due to the relative scale-independence of $S_3$. 
\footnote{Bernardeau (1995) pointed out a systematic correction to the 
way the $S_J$ predictions should be projected.  
After taking into account the correct selection function 
and the uncertainties in the APM shape of $P(k)$, this effect 
is less significant that claimed 
by Bernardeau, and the original interpretation is still valid,
in a greement with the results for $q_3$ presented here.}
On the other hand, Fry (1994) used the projected bispectrum 
from the Lick catalog to infer $b_1 \simeq 3$; 
however, in order to 
extract a statistically significant $Q_g$ from the catalog,
an average over scales including those beyond the weakly non-linear 
regime was required. As we will see below, the 
configuration-dependence of $Q_{g}$ on large scales in the APM 
catalog  
is quite close to that expected in PT, suggesting 
that $b_1$ is of order unity for these galaxies. 
The simple model above undoubtedly does not capture the full 
complexity of biasing (e.g., Mo \& White 1996, 
Blanton et al. 1998, Dekel \& Lahav 1998, Sheth \& Lemson 1998), but 
it provides a convenient framework that 
is well matched to the quality of the current data.

In a projected catalog with radial selection function $\phi(x)$ (normalized 
such that $\int dx x^2 \phi(x) =1$),  
the galaxy angular two- and three-point functions at small angular 
separations ($\theta_{ij} \ll 1$) are given by (Peebles \& 
Groth 1975, Peebles 1980) 
\beq
w_g(\theta)=2 \int_0^\infty dx {x^4 \phi^2(x) F^{-2}(x) }
\int_0^\infty du ~\xi_g(r;z)
\label{wt}
\eeq
\beq
z_g(\theta_{12}, \theta_{13}, \theta_{23})= \int_0^\infty dx {x^6 \phi^3(x)
F^{-3}(x)}\int_{-\infty}^\infty \int_{-\infty}^\infty du dv 
~\zeta_g(r_{12}, r_{13}, r_{23}; z) ~,
\label{zt}
\eeq
where $\theta_{12}, \theta_{13}$, and $\theta_{23}$ are the sides of 
a triangle projected on the sky.
We assume the proper separations $r_{ij}$ are small compared to 
the mean depth of the sample; in this 
case,  
\beq
r_{ij} = {1\over 1+z}\left[(f_{ij}/F)^2+x^2 \theta_{ij}^2\right]^{1/2}~,
\eeq
with $f_{12}=u$, $f_{13}=v$, and $f_{23}=u-v$. Here, $x$ is the 
comoving radial (coordinate) 
distance to redshift $z$, and $F(x,\Omega)$ is a geometrical 
factor which relates proper and coordinate distance intervals, 
$F(x)=[1-(H_0x/c)^2(\Omega-1)]^{1/2}$.

The radial selection function for the APM Galaxy Survey can be 
approximated by (Gazta\~naga \& Baugh 1998)
\beq
\phi(x)_{APM} =C x^{-b} \exp^{-x^2/D^2}~,
\label{selfn}
\eeq
with $b=0.1$ and $D=335 h^{-1}$ Mpc.
At this depth, to an accuracy of better than a few percent 
we can approximate $F$ by its Einstein-de Sitter 
value $F=1$. The projected hierarchical amplitude is defined by analogy with 
eqn.(\ref{Q3def}), 
\beq
q_3(\theta_{12}, \theta_{13}, \theta_{23}) \equiv {
z_g(\theta_{12}, \theta_{13}, \theta_{23}) \over
w_g(\theta_{12})w_g(\theta_{13})+w_g(\theta_{12})w_g(\theta_{23})+
w_g(\theta_{13})w_g(\theta_{23})}~.
\label{Qp}
\eeq
In projecting eqns.(\ref{xidef})
and (\ref{zeta1}) in PT we assume leading-order 
perturbative growth for the redshift-evolution of $\xi$ and $\zeta$ for 
$\Omega_m=1$; in 
this case, both $Q$ and $q_3$ are independent of the power spectrum 
normalization (e.g., independent of $\sigma_8$).

For the analysis, we use the equal-area projection pixel map of the 
APM survey, with an area roughly $120^o \times 60^o$, 
containing $N_{gal} \simeq 1.3\times 10^6$ galaxies brighter 
than $b_j = 20$ and fainter than $b_j = 17$. 
Each pixel has an area $(\Theta_p)^2\simeq (0.06)^2$ sq. deg., 
and the mean galaxy count per pixel is $\langle N \rangle =N_{gal}/N_{pix} 
=0.97$. The estimator of the 
density fluctuation amplitude at the $i$th pixel is 
${\hat\delta}_i = (N_i/\langle N \rangle)-1$, where $N_i$ is the galaxy 
count in that pixel. The estimator for the galaxy angular 
two-point function is then (Peebles \& Groth 1975) 
\beq
{\hat w}(\theta) = {1\over N^{(2)}_\theta}\sum_{i,j} \delta_i 
\delta_j W_{ij}(\theta)~,
\label{2est}
\eeq
where $N^{(2)}_\theta = \sum_{i,j} W_{ij}(\theta)$ is the number of 
pairs of pixels at separation $\theta$ in the survey region, and the 
angular window function 
$W_{ij}(\theta)=1$ if pixels $i$ and $j$ are separated by 
$|{\vec \theta}_i - {\vec \theta}_j| = \theta \pm \Theta_p$,  
and 0 otherwise. The reduced 
angular three-point function is estimated as
\beq
{\hat z}(\theta_{12}, \theta_{13}, \theta_{23}) = {1\over N^{(3)}_\theta}
\sum_{i,j,k} \delta_i \delta_j \delta_k W_{ijk}
(\theta_{12}, \theta_{13}, \theta_{23})~,
\label{3est}
\eeq
where $N^{(3)}_\theta = \sum_{i,j,k} W_{ijk}$ is the count of triplets 
of pixels with angular separations $\theta_{12}, \theta_{13}, \theta_{23}$ 
in the survey region, and $W_{ijk}=1$ for pixel triplets with that 
angular configuration and 0 otherwise. We note that, in the limit of 
counts of pairs and triplets 
of objects, these estimators are equivalent to the minimum variance 
estimators of Landy \& Szalay (1993) and Szapudi \& 
Szalay (1998). We employ these estimators for angular 
separations $\geq 0.5$ deg, at least an order of magnitude larger than 
the pixel scale; in this limit, finite pixel-size corrections should 
be less than a few percent and have been neglected. 

To test the validity of PT on the scales of interest, to verify  
the algorithm for measuring the angular three-point function, 
and to check projection and finite sampling (and boundary) effects in 
the APM survey, we use the simulated APM maps of 
Gazta\~naga \& Baugh 1998. These are created from N-body simulations 
of the SCDM (with $\sigma_8=1$) and APM-like ($\sigma_8=0.8$) 
models, with box size of $600 ~h^{-1}$ Mpc and 
$128^3$ particles, both with $\Omega_m = 1$.  From each N-body 
realization we make 5 APM-like maps from 5 different observers; the 
observers are spaced sufficiently far apart compared to $D$ 
that the `galaxies' they 
observe do not appear to be strongly correlated (Gazta\~naga \& 
Bernardeau 1998).
As the simulation is done in a periodic box, we replicate the box to cover
the full radial extent of the APM (around 1800 $h^{-1}$ Mpc, at which 
distance the expected number of galaxies is smaller than unity).
To account for possible boundary effects, we employ the 
APM angular survey mask, including plate shapes and holes. 
We have also made larger angular maps without the APM mask and find
no significant differences from the results shown below, 
from which we conclude that the APM mask
does not affect the estimations of the 2 or 3-point functions on 
the scales under study here. The projected simulations have $\sim 30\%$ lower
mean surface density of galaxies than the APM. We have diluted the APM pixel 
maps by this amount (by randomly sampling the 
galaxies) and found that this dilution has a negligible effect on the 
clustering properties within the errors.

To estimate the projected 3-point function we need to find all triplets in the 
pixel maps which have a given triangle configuration and repeat the process 
for all configurations. 
Consider a triplet of pixels with labels 1,2,3 on the sky. Let 
$\theta_{12}$ and $\theta_{13}$ be the angular separations between the 
corresponding pairs of pixels and $\alpha$ the interior angle between 
these two triangle sides. One can characterize the 
configuration-dependence of the three-point function by 
studying the behavior of $q_3(\alpha)$ for fixed
$\theta_{12}$ and $\theta_{13}$.
Our algorithm for counting such triplets is as follows. 
For each pixel in the map, 
we find all pixels that lie in two concentric annuli of 
radii $\theta_{12}\pm 0.5$ and $\theta_{13}\pm 0.5$ about it, where
$\theta_{ij}$ is now measured in units of $\Theta_p$. We count 
all pairs of pixels between the two annuli; this  
requires $(m_{12} \times m_{13})/2$ operations, where 
$m_{ij}= 2 \pi \theta_{ij}$. As seen from the central pixel, 
each pair has an angular separation $\alpha$, and the results are 
binned in $\alpha$. We repeat this 
procedure for each pixel in the map, building the estimators (\ref{2est}) 
and (\ref{3est}). Thus, to find all triplets with separations 
$\theta_{12}$ and $\theta_{13}$ requires only $N_{pix}(m_{12} \times m_{13})/2$
operations; for $\theta_{12}= \theta_{13}=1$ deg, this computation 
only takes a few hours of CPU time on a modest workstation. By 
contrast, a naive ${\cal O}(N_{pix}^3)$ operation with $N_{pix}\simeq 10^6$ 
would be quite a lengthy computational task given 
current computer power.

\begin{figure}[t!]
\centering
\centerline{\epsfxsize=18truecm\epsfysize=18truecm\epsfbox{
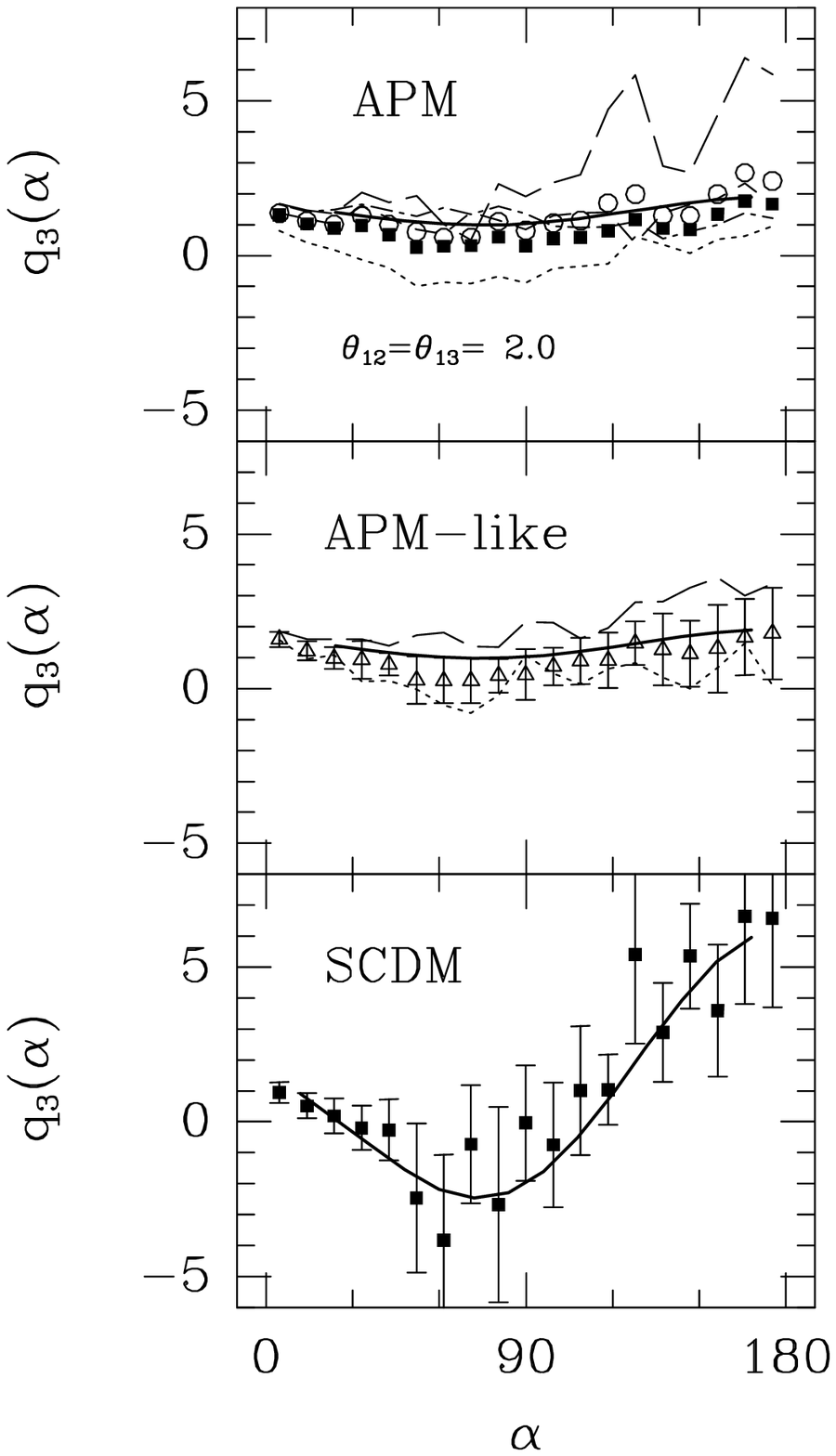
}}

\caption[junk]{The two lower panels show perturbation theory predictions 
(solid curves) and N-body results (points with sampling errors) 
for the projected 3-point amplitude $q_3(\alpha)$ at fixed 
$\theta_{12}=\theta_{13}=2$ deg for a survey 
with the APM selection function. N-body results are the mean 
from 10 observers. Middle panel corresponds to 
the initial APM-like power spectrum of Eqn.(\ref{Pkapm}) and lower 
panel to the SCDM model. Additional curves in the middle panel 
show results for two different observers. The top panel compares the 
PT prediction for the APM-like model 
with the real APM measurements (closed squares and open circles 
correspond to the full APM map and to the mean of 4 zones; other 
curves in the top panel show results for each of the zones).} 

\label{q3nbodyr20}
\end{figure}

The two lower panels in Fig. \ref{q3nbodyr20} show results 
for $q_3(\alpha)$ for the 
SCDM and APM-like models, for $\theta_{12}=\theta_{13} = 2$ degrees, 
projected at the depth of the APM survey. The configuration-dependence 
of the hierarchical three-point amplitude is seen to be quite sensitive to the 
shape of the power spectrum. Both the shape and amplitude of $q_3(\alpha)$
predicted by PT (solid curves) are 
reproduced by the N-body results (points) even on these 
moderately small scales (at the mean depth of the APM, 2 deg corresponds 
to $\simeq 14$ h$^{-1}$ Mpc). Part 
of this agreement traces to the fact that $q_3$ involves a weighted 
sum over spatial 3-point configurations covering a range of scales; 
due to the shape of the power spectrum, $Q$ increases on large 
scales, so these configurations (which are further in the perturbative regime) 
are more heavily weighted in projection. The error bars on the simulation 
results are estimated from the variance between 10 maps (5 observers 
each in 2 N-body realizations), assuming 
they are independent, and correspond to the 1-$\sigma$ interval of confidence
for a single observer (i.e., they are not divided by $\sqrt{10-1}$).

\begin{figure}[t!]
\centering
\centerline{\epsfxsize=18truecm\epsfysize=18truecm\epsfbox{
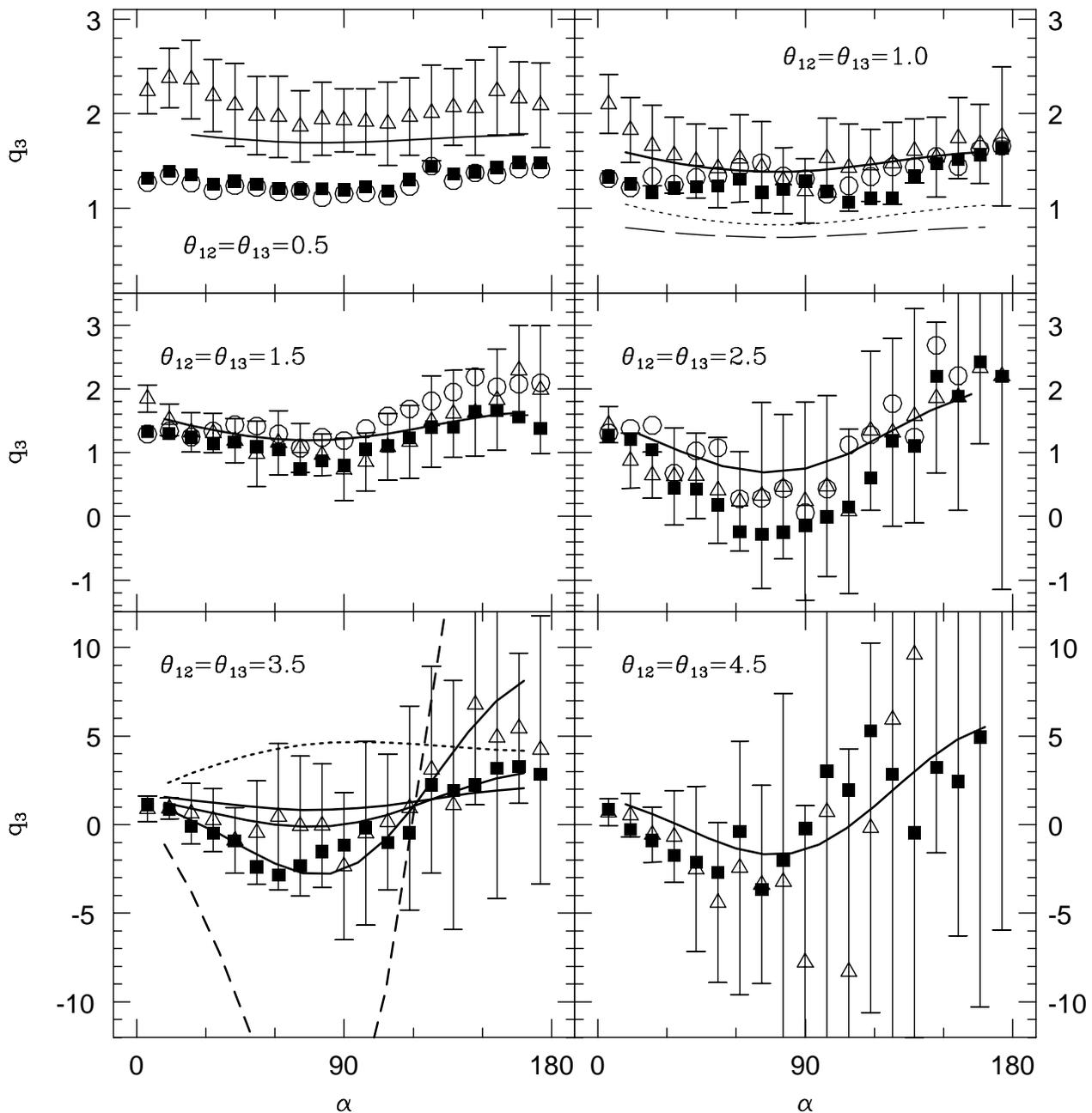
}}

\caption[junk]{The projected 3-point amplitude $q_3$ in PT (solid curves) 
and N-body results (open triangles with errorbars) for the APM-like power 
spectrum are compared with $q_3$ measured in 
the APM survey (closed squares and open circles, with same meanings 
as in Fig. 1). Each panel shows the amplitude at different
$\theta_{12}=\theta_{13}$. In upper right panel, dotted and 
dashed curves correspond to PT predictions with $b_1=1, b_2=-0.5$ 
and $b_1=2, b_2=0$, respectively. In the lower left panel, upper 
and lower solid curves conservatively bracket the uncertainties in 
the inferred APM-like power spectrum, long-dashed curve corresponds to SCDM, 
and the dotted curve shows the leading-order prediction for the 
$\chi^2$ non-Gaussian model.}
\label{q3panel}
\end{figure}

The top panel of Fig. \ref{q3nbodyr20} and all those of 
Fig. \ref{q3panel} show the measurements of $q_3(\alpha)$ in the  
APM survey itself at $\theta_{12}=\theta_{13}=0.5-4.5$ degrees.
Closed squares correspond to estimations in the full APM map. 
The values at $\alpha=0$ are in agreement with
the cumulant correlators, $c_{12} \equiv q(0)$, 
estimated (with 4x4 bigger pixels) by Szapudi \& Szalay (1999).
The mean values are comparable to the values of $s_3/3$ in the APM 
(Gaztanaga 1994, Szapudi et al. 1995). Also notice that at the
scales considered here, $\geq 0.5$ deg, the values of $s_3$ in the APM and 
the EDSGC (Szapudi, Meiksin, \& Nichol 1996) are very 
similar (Szapudi \& Gaztanaga 1998).

The APM results are compared with the values of $q_3$ for
the APM-like spectrum in PT (solid curves) and in simulations 
(open triangles with errorbars).
As the APM-like model has, by construction, 
the same $w(\theta)$ as the real APM map, we assume that the sampling 
errors should be similar in the APM and in the simulations.  
This might not be true on the largest scales, however, 
where systematics in both the
APM survey and the simulations are more important 
(e.g., the simulation might be affected 
by the periodic boundaries, and the power in the simulation and in the 
survey may differ on the largest scales). 
At scales $\theta \ga 1$ deg, the 
agreement between the APM-like model and the APM survey is quite good; 
this corresponds roughly to physical scales 
$r \ga 7$ h$^{-1}$ Mpc, not far from 
the non-linear scale (where $\xi_2 \simeq 1$).
At $4.5$ deg, the APM agrees better with the PT prediction than 
with the simulation results, which show large variance; 
this could be an indication that the 600 h$^{-1}$ Mpc simulation box 
is not big enough, leading to larger sampling errors than the
APM itself. We also note that the SCDM model clearly disagrees with 
the APM data for $q_3$;  
this can be seen at $\theta_{12}=\theta_{13}=2$ deg 
(Fig.\ref{q3nbodyr20}) and for $\theta_{12}=\theta_{13}=3.5$ deg 
(dashed curve in lower-left panel of  Fig. \ref{q3panel}). This 
conclusion is independent of the power spectrum normalization.

At smaller angles, $\theta \la 1$ deg, $q_3$ in the simulations is 
larger than in either the real APM or PT (top-left panel in
 Fig. \ref{q3panel}). The discrepancy between simulations and PT on these 
relatively small scales is clearly due to non-linear evolution. The 
interpretation of the discrepancy with the 
real APM is less clear: a number of assumptions underlying the simulations 
could affect the final results at these non-linear scales. 
For example, systematic uncertainties in the APM selection function 
or a linear bias would lead one to infer a different linear 
APM-like power spectrum from the 
$w(\theta)$ data. Also, a model with low $\Omega_m$ would undergo 
less non-linear evolution, which might give a better match to the 
APM results for $q_3$; this could provide an interesting test for 
a low-density universe. This discrepancy 
at non-linear scales is similar to the one found in Baugh \& 
Gazta\~naga (1996), where the real APM values of $S_J$ were 
closer to the PT predictions than to simulations. Other possible 
contributions to this effect include non-linear bias and 
non-linear projection effects (Gazta\~naga \& Bernardeau 1998).

The open circles in each figure  
show the mean of the estimations of $q_3$ in 4 disjoint subsamples of the APM
survey (equally spaced in right ascension, as in Baugh \& Efstathiou 1994).  
For illustration, the values of $q_3$ for each of the 4 zones are shown 
in the top panel of Figure \ref{q3nbodyr20}: the 
dotted, short-dashed, long-dashed, and dot-dashed curves 
correspond to zones of increasing RA 
(the middle two of these correspond to relatively lower galactic lattitude).
These estimations of $q_3$
are subject to larger finite-volume effects, because each zone is only 
1/4 the size 
of the full APM. Because the zones cover a range of galactic latitude, 
a number of the systematic errors in the APM catalog 
(star-galaxy separation, obscuration by the galaxy, plate matching errors)
might be expected to vary from zone to zone. We find no evidence for 
such systematic variation in $q_3$: the individual
zone values are compatible with the full survey
within the (sampling) errors in the simulations 
(compare the top and middle panels in Figure \ref{q3nbodyr20}). 
On larger scales, $\theta \ga 3$ deg, the individual zone amplitudes
exhibit large variance, and boundary effects come into play.

On all scales, we find large covariance between the errors of 
$q_3$: the data points at different $\alpha$ are strongly correlated. 
This is illustrated in the middle panel of Fig. 
\ref{q3nbodyr20}: the dotted and continuous curves correspond to results for 
2 of the 10 observers. Sampling or finite-volume effects are seen to 
produce a systematic vertical shift in the curve rather than a scatter around 
some mean value. A similar trend is found in the real APM catalog 
(compare the 4 curves in the top panel of  Fig. \ref{q3nbodyr20}). 
This covariance can be studied
analytically (Hui, Scoccimarro, Frieman, \& Gaztanaga, in preparation) 
and comes from fluctuations on scales comparable to the sample size.
Ratio bias and integral-constraint bias (e.g., Hui \& Gaztanaga 1999) could 
also be important.

Other possible sources of systematic discrepancies between the model 
predictions and the APM results include 
the shape of the APM radial selection function, the evolution of 
clustering, and the shape of the linear power spectrum.
We find that the first two effects introduce differences smaller than $10\%$ 
in the amplitude of $q_3$ 
(in agreement with Gaztanaga 1995), which are not significant given the errors.
The uncertainty in the shape of the linear $P(k)$ is more important, and 
as mentioned above is critical for the interpretation of $q_3$ 
at the smallest angles. Nevertheless, 
at large scales ($\theta > 1$ deg) these uncertainties appear to be within the
errors (when we take into account that the errors are strongly correlated).
This is illustrated in the bottom-left panel of  Fig. \ref{q3panel}---the 
two solid curves bracketing the APM-like results 
show the PT predictions for two power spectra which conservatively 
bracket the uncertainties in the linear spectrum inferred from 
the APM $w(\theta)$ 
(see Fig.13 of Gaztanaga \& Baugh 1998): 
$P(k) \sim k^{a}/[1+(k/0.06)^3]$, with $a=0.2$ and $1.2$. The model 
with $a=1.2$ (the lower solid curve at small $\alpha$)   
appears to give a better match to the $q_3$ results in the APM 
than the central APM-like model of Eq.[\ref{Pkapm}]; for reference, 
a CDM model with $\Gamma=0.3$ gives a nearly identical value for $q_3(\alpha)$ 
on these scales. Thus, although the APM results for 
$q_3$ generally fall below the PT predictions 
on angular scales $\theta \ga 2$ deg, they are consistent 
within the sampling errors and given the uncertainties in the 
shape of the linear power spectrum inferred from the APM $w(\theta)$.




To place accurate constraints upon bias models and initially 
non-Gaussian fluctuations, we must quantitatively model the 
covariance between the estimates of $q_3$; this will be done elsewhere, 
but we can nevertheless get a qualitative sense of the limits here. 
We expect the strongest constraints to come from
intermediate scales, $\theta \simeq 1-2.5$ deg, where both the 
sampling errors and the non-linearities are small. 
The upper right panel of Fig. \ref{q3panel} shows the PT 
predictions for the APM-like model with linear bias parameter 
$b_1=2$ (dashed curve) and a non-linear bias model 
with $b_1=1$, $b_2=-0.5$ (dotted curve). 
Even if the errors are $100\%$ correlated, 
these models are clearly ruled out by the APM data; we 
conservatively conclude that $b_1 \la 1.5$ is required for a 
simple linear bias model to fit the APM data.
Note that a model with 
$b_1=1.5$ and $b_2=0.5$ would have roughly the correct 
amplitude for $q_3$, but its shape would be flatter than the 
data, especially at larger $\theta$.   

As a simple example of a non-Gaussian model, the dotted curve 
in the lower-left panel of  Fig. \ref{q3panel}
shows the leading-order prediction for the $\chi^2$ isocurvature 
model (Peebles 1997, 1998a, 1998b, 
Antoniadis, etal. 1997, Linde \& Mukhanov 1997, White 1998) 
with the APM-like spectrum. In this model, the initial density
field is the square of a Gaussian random field, and 
the leading-order 3-point function 
is simply $\zeta = 2[2 \xi(x_{12})\xi(x_{13})\xi(x_{23})]^{1/2}$.
Clearly, the projected three-point function for this model is 
substantially larger than that of the corresponding Gaussian model for 
intermediate $\alpha$; both the amplitude and shape are discrepant  
with the APM data. To make 
this comparison precise, the non-linear corrections for this 
model should be self-consistently included; however, these 
corrections are expected to increase rather than reduce the 
$q_3$ amplitude, likely making the disagreement worse. 
In general, if we assume no biasing, the initial non-gaussianities
are restricted to  $|q_3| \la 0.5$.
             
At angular scales $\theta \ga 1$ deg, corresponding to physical 
scales for which  $\xi \la 1$, the 
agreement between PT and the APM survey for the angular 
three-point amplitude $q_3$ is quite good, implying 
that APM galaxies are not significantly biased on these scales and 
that their spatial distribution is consistent with 
non-linear  evolution from Gaussian 
initial conditions. This substantiates and extends the conclusions of 
Gazta\~naga (1994,1995) and Gazta\~naga \& Frieman (1994). 

\acknowledgements We thank  A. Buchalter, A. Jaffe, and M. 
Kamionkowski, who have recently carried out an independent 
computation of the angular three-point function in PT, for 
disucssions, as well as 
L. Hui, R. Juszkiewicz, and R. Scoccimarro. EG would like to thank 
G.Dalton, G.Efstathiou and the Astrophysics group
at Oxford, were this work started.
This research was supported in part by the DOE 
and by NASA grant NAG5-7092 at Fermilab and by a NATO 
Collaborative Research Grants Programme CRG970144 between IEEC
and Fermilab. EG also acknowledges support by 
IEEC/CSIC and by DGES(MEC) (Spain), project PB96-0925.

%
%


\clearpage



\end{document}